\documentclass[global,twocolumn]{svjour}
\usepackage{epsf}
\journalname{Applied Physics A}
\begin{document}
\title{Critical Exponents for the Ferromagnetism in
Colossal Magnetoresistance Manganites
}
\author{Nobuo Furukawa\inst{1} \and Yukitoshi Motome\inst{2}
}                     
\offprints{Nobuo Furukawa}          
\institute{Department of Physics, Aoyama Gakuin University, Setagaya, Tokyo, Japan
\and Institute of Materials Science, University of Tsukuba, Tsukuba, Ibaraki, Japan}
\date{Received: date / Revised version: date}
%
\maketitle
\begin{abstract}
Critical phenomena of ferromagnetic transition
in colossal magnetoresistance manganites are theoretically studied.
Concerning the critical exponents for this transition,
there still remains controversy among experimental results.
In order to clarify intrinsic physics of the manganites
through a comparison with theoretical prediction,
we investigate the critical phenomena of double-exchange models
by using finite-size scaling analysis on unbiased numerical results.
As a result, we show that the critical exponents of
the ferromagnetic transition of the three-dimensional double-exchange model
is consistent with those of the Heisenberg model,
but are distinct from the mean-field one.
\end{abstract}
%
\section{Introduction}
\label{Sec:Introduction}

Ferromagnetic transition in doped manganese oxides has revived interests
since rediscovery of colossal magnetoresistance phenomena.
\cite{Ramirez1997,Furukawa1999}
Recent refinement of experimental technique and
improvement of sample quality make it possible
to discuss critical phenomena of this transition in detail.
Particularly, for the purpose of capturing intrinsic physics
in these materials and promoting a better understanding of
the colossal magnetoresistance phenomena,
the critical exponents and the universality class
have been investigated by various experimental methods.
However, the estimates of the exponents are scattered and have not converged yet.
For instance, the neutron scattering gives Heisenberg-like universality class,
\cite{Martin1996,Vasiliu-Doloc1998}
and some d.c. and r.f. magnetization measurements give mean-field-like one.
\cite{Mohan1998,Schwartz2000}
It is strongly desired to solve this controversy.

The purpose of this work is to shed light on this problem
from a theoretical viewpoint.
We study the universality class of the ferromagnetic transition
due to double-exchange (DE) mechanism, which is believed to be
a major driving force to cause the metallic ferromagnetism in doped manganites.
We compare the results with experimental data, and
discuss the origin of the experimental controversy.

The DE mechanism itself was proposed more than half-century before,
\cite{Zener1951}
and has been studied intensively.
Nevertheless, the thermodynamics has not been fully understood thus far.
Large spin fluctuations and the strong interplay
between charge and spin degrees of freedom make it difficult
to describe finite-temperature properties quantitatively,
especially near the critical point of the ferromagnetic transition.
In order to give a quantitative description of the critical phenomena,
it is indispensable to take account of the large fluctuations.
Approximational methods like the mean-field approximation are known to fail.
\cite{Furukawa1999}

As one of promising candidates to study this problem, in this paper,
we employ a numerical calculation which includes the fluctuations properly
and is free from uncontrolled approximations.
Recent development of algorithm and computational environment
enables us to perform large-scale unbiased calculations and
to describe the finite-temperature properties in the thermodynamic limit.
We analyze the numerical results which are obtained
by a newly developed Monte Carlo (MC) method
in the next section \ref{Sec:Results}.
In Sec. \ref{Sec:Discussion}, we discuss the results and compare them
with experimental results.
Section~\ref{Sec:Summary} is devoted to summary.

\section{Numerical Results}
\label{Sec:Results}
We develop and employ a new MC algorithm,
which is called the moment-expansion MC method.
\cite{Motome1999}
By using the moment expansion of the density of states
which is efficiently performed on parallel computers,
this new method reduces the computational time remarkably
compared to the conventional MC technique.

We study DE models with a single conduction band
(which stands for Mn $e_{g}$ bands)
which couples to localized spins (in Mn $t_{2g}$ localized levels).
The Hamiltonian is given by
\begin{equation}
{\cal H} = -t \sum_{\langle i,j \rangle, \sigma}
c_{i \sigma}^{\dagger} c_{j \sigma}
- J_{\rm H} \sum_{i} \vec{\sigma}_{i} \cdot \vec{S}_{i},
\label{eq:H_DE}
\end{equation}
where the first term describes the electron transfer
between nearest-neighbor sites, and
the second one is for the Hund's-rule coupling ($J_{\rm H} > 0$)
between the conduction electron spin $\vec{\sigma}$ and
the localized spin $\vec{S}$.
For simplicity, hereafter, we consider the strong coupling limit,
$J_{\rm H} \rightarrow \infty$,
and the classical localized spin, $|\vec{S}| \rightarrow \infty$.
The following calculations are done at quarter-filling.

\begin{figure}
\epsfxsize=6.5cm
\centerline{\epsfbox{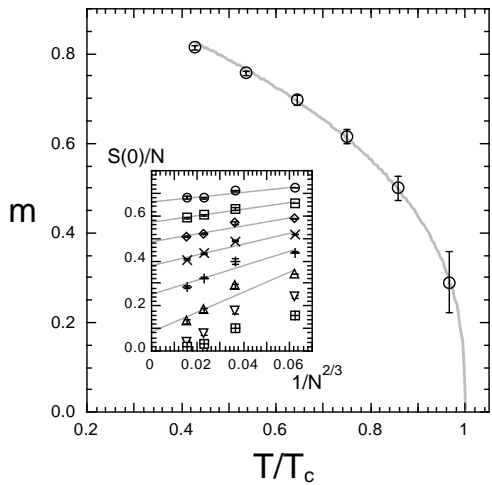}}
\caption{
Temperature dependence of the magnetization in the thermodynamic limit.
The curve in the figure is the least-squares-fit to
$m \propto (T_{\rm c} - T)^{\beta}$
with the Heisenberg exponent $\beta=0.365$.
The inset shows the system-size extrapolation of the spin structure factor.
}
\label{fig:mvsT}
\end{figure}

\begin{figure}
\epsfxsize=6.5cm
\centerline{\epsfbox{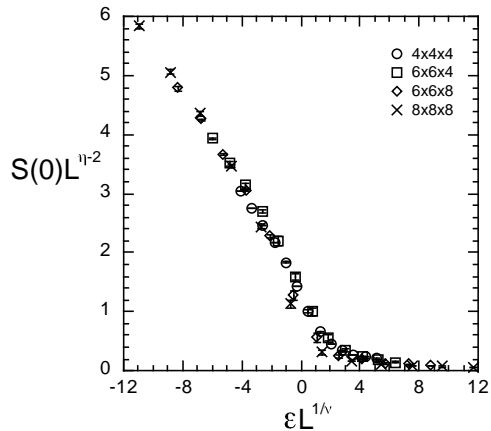}}
\caption{
Finite-size scaling plot
for the spin structure factor in the three-dimensional model.
Critical exponents in the Heisenberg universality class are assumed.
}
\label{fig:scaling3D}
\end{figure}

First, we study a three-dimensional DE model
with Heisenberg spin symmetry, which is realistic
for comparison with experiments.
\cite{Motome2000}
We analyze the MC data by assuming the universality class
of Heisenberg models with short-range interactions.
Figure~\ref{fig:mvsT} shows the temperature dependence of the magnetization
in the thermodynamic limit.
The data are obtained by system-size extrapolation of 
the spin structure factor as shown in the inset.
The magnetization data scale well with
the fitting $m \propto (T_{\rm c} - T)^\beta$
with the Heisenberg exponent $\beta = 0.365$.
Figure~\ref{fig:scaling3D} shows the finite-size scaling
for the spin structure factor of the ferromagnetic component $S(0)$.
The scaling relation is given by
\begin{equation}
S(0)L^{\eta-2} = f( \varepsilon L^{1/\nu} ),
\label{eq:scaling}
\end{equation}
where $L$ is the linear dimension of the finite-size clusters,
$f$ is the universal scaling function and $\varepsilon = (T-T_{\rm c})/T_{\rm c}$.
In Fig.~\ref{fig:scaling3D}, we assume the Heisenberg exponents,
$\nu = 0.705$ and $\eta = 0.034$.
The MC data for different $L$ and $T$ appear to follow a universal function.
We have tried the mean-field exponents $\beta = \nu = 0.5$ and $\eta = 0$ also,
and we confirm that the Heisenberg exponents give better scaling fits
than the mean-field ones.

\begin{figure}
\epsfxsize=6.5cm
\centerline{\epsfbox{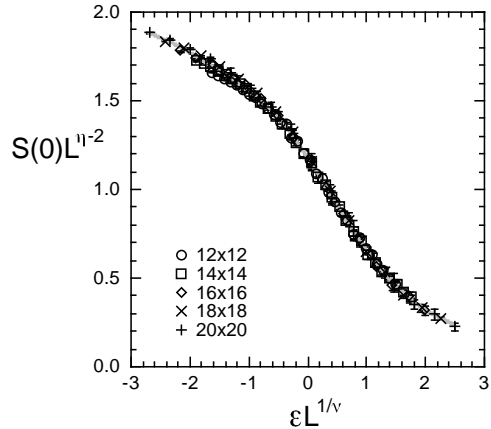}}
\caption{
Finite-size scaling plot of
the spin structure factor in the two-dimensional model.
Critical exponents of Ising universality class are assumed.
}
\label{fig:scaling2D}
\end{figure}

Next, in order to perform more intensive and precise MC samplings
in large-size systems,
we study a simplified model of eq.~(\ref{eq:H_DE}), that is,
a two-dimensional DE model with Ising spin symmetry.
\cite{Motome2001}
The localized spins are Ising variables which take $S = \pm 1$.
The model is much simplified, however,
it still include the essence of the DE mechanism, that is,
the kinetics of electrons is strongly correlated with spin degrees of freedom
and energetically favors ferromagnetism.
We analyze the MC data on spin structure factor
by finite-size scaling analysis as in Fig.~\ref{fig:scaling3D}.
Figure~\ref{fig:scaling2D} shows the result
on the assumption of the exponents for the Ising spin system
in two dimensions, $\nu = 1$ and $\eta = 0.25$.
All the MC data are scaled well with a universal function.
In this case also, we confirm that the mean-field exponents,
$\nu = 0.5$ and $\eta = 0$, do not give a satisfactory scaling.

This obvious result in the two-dimensional model strongly supports
the previous conclusion in three dimensions, i.e.,
the critical exponents in the DE model are consistent with
the universality class of the Heisenberg models with short-range interactions.
The estimates of the exponents are distinct from the mean-field ones.
These results are discussed and compared with experimental results
in the next section.

\section{Discussion}
\label{Sec:Discussion}

The DE interaction has a distinguishable property
compared to ordinary exchange interactions in spin systems.
Effective ferromagnetic interaction is induced by the kinetics of electrons
which favor extended states with ferromagnetic spin background
to gain the kinetic energy.
If we integrate out the electron degrees of freedom
to describe the action as a function of spin configurations,
it is necessary to introduce effective long-range two-spin interactions as well as
multiple-spin interactions which depend on sizes and shapes of
ferromagnetic domain structure.
The range of the interaction is determined
to minimize the total free energy for charge and spin degrees of freedom.
As the system approaches the critical point,
the magnetic domain structure fluctuates strongly.
Thus, it is highly nontrivial how the DE interaction is renormalized,
whether it is renormalized to a short-range one, or
the long-range and the multiple-spin interactions become relevant
to cause the mean-field-like transition through suppression of fluctuations.

The results in the previous section indicate that
the universality class of the ferromagnetic transition in the DE model
is consistent with that of models with short-range interactions.
It is distinct from the mean-field universality class.
Thus, we conclude that the DE interaction does not become
effectively long-range, but is renormalized to be short-range
near the critical point.

As mentioned in Sec.~\ref{Sec:Introduction},
experimental estimates of the critical exponents are widely scattered
including the Heisenberg-like (short-range) and the mean-field-like (long-range) ones.
Our numerical results indicate that
the short-range Heisenberg universality class should be observed in real materials
when the ferromagnetic transition is ascribed to the DE mechanism as a major origin.

In order to solve the experimental controversy,
the following points should be considered carefully.
The first point is sample quality.
For precise estimates of the critical exponents,
it is crucial to determine the critical temperature $T_{\rm c}$ precisely.
Sharpness of the transition as well as the value of $T_{\rm c}$
depends strongly on homogeneity of samples.
The second point is the range of the critical region.
The exponents may depend on the region where they are estimated,
particularly when different interactions compete with each other.
Generally, it is difficult to determine true critical region, but
the dependence of the exponents on the critical range should be tested.
The third point is that, in real materials,
there may be other elements which are neglected
in the simple DE model (\ref{eq:H_DE}), for instance,
orbital degrees of freedom, Coulomb repulsion between conduction electrons, and
electron-phonon coupling.

We comment on the final point.
We believe that these additional elements neglected in model (\ref{eq:H_DE})
do not alter the short-range universality class of the ferromagnetic transition
because they will not suppress the fluctuations.
The spin fluctuations may become more strong through these additional channels.
At the same time, the electrons become more incoherent
by these additional elements.
This makes the effective magnetic interaction more short-range
since the incoherence shortens the range of the effective electron transfer.
These considerations support our conclusion that
the short-range Heisenberg universality class will be found in experiments
if the transition is purely magnetic.

\section{Summary}
\label{Sec:Summary}

We have investigated the critical phenomena of the ferromagnetic transition
by the double-exchange mechanism
in order to discuss the experimental controversy
on the estimates of the critical exponents.
We have estimated the critical exponents
by using the finite-size scaling analysis on the Monte Carlo data.
The obtained exponents suggest that the universality class of
the three-dimensional double-exchange model
is the same as that of the Heisenberg models with short-range interactions.
The double-exchange interaction is renormalized to be short-range
when the system approaches the critical point.
This theoretical result provides a useful reference
to consider the experimental controversy.
We have discussed possible origins of the discrepancy among experiments.
Although our conclusion is favorably compared with the neutron scattering data,
further experimental studies are desired to settle this controversy.
A conclusive determination of the critical exponents and the universality class
will help us to understand the intrinsic physics of doped manganites and
of the colossal magnetoresistance phenomena.

\end{document}